\documentclass[11pt,a4wide]{article}
\usepackage{newlfont}
\usepackage{amsmath}
\usepackage{graphics}
\usepackage{float}
\usepackage{graphicx}
\usepackage{epsfig}
\usepackage{multirow}
\begin{document}

\title{Potential Model for $\Sigma_{u}^{-}$ Hybrid Meson State}
\author{Nosheen Akbar\thanks{e mail: nosheenakbar@cuilahore.edu.pk,noshinakbar@yahoo.com}\quad and Saba Noor\thanks{e mail: sabanoor87@gmail.com} \\
\textit{$\ast$Department of Physics, COMSATS University Islamabad, Lahore Campus,}\\
{Lahore(54000), Pakistan.}\\
\textit{$\dag$Centre For High Energy Physics, University of the Punjab, Lahore(54590), Pakistan.}}
\date{}
\maketitle

\begin{abstract}
In this paper, lattice simulations are used to propose a potential model for gluonic excited $\Sigma^-_u$ states of bottomonium meson .
This proposed model is used to calculate radial wave functions, masses and radii of $\Sigma_u^-$ bottomonium hybrid mesons.
Here, gluonic field between a quark and an antiquark is treated as in the Born-Oppenheimer expansion, and Schr$\ddot{o}$dinger equation is numerically solved employing shooting method. Results of calculated masses for $\Sigma^-_{u}$ state are in quite good agreement with the lattice simulations.
  \end{abstract}


\section*{I. INTRODUCTION}

Static quark potential models play important role in the understanding of Quantum chromodynamics. A hybrid static potential is defined as a potential of a static quark-antiquark pair with the gluonic field in the excited states. These hybrid static potentials for different states of mesons are computed in refs. \cite{morning}\cite{Bali}\cite{2019}\cite{adam}\cite{morningstar03}. Hybrid static potentials are characterized by quantum numbers, $\Lambda$, $\eta$, and $\epsilon$, where $\Lambda$ is the projection of the total angular momentum of gluons and for $\Lambda = 0, \pm1, \pm2, \pm3, .... $, meson states are represented as $\Sigma,\Pi,\Delta$ and so on \cite{morning}. $\eta$ is the combination of parity and charge and for $\eta = P \circ C = +,-$, states are labelled by sub-script $g,u$ \cite{morning}. $\epsilon$ is the eigen value corresponding to the operator $P$ and is equal to $+,-$. Parity and charge for hybrid static potentials are defined as \cite{morning}
\begin{equation}
P=\epsilon (-1)^{L+\Lambda +1},C= \epsilon \eta (-1)^{L+\Lambda +S},
\label{cp}
\end{equation}
The low-lying static potential states are labelled as $\Sigma_g^{+}, \Sigma_g^{-},\Sigma_u^{+},\Sigma_u^{-}, \Pi_{g}, \Pi_{u}, \Delta_{g}, \Delta_{u}$ and so on\cite{morning}. $\Sigma_g^{+}$ is the low-lying potential state with ground state gluonic field and is approximated by a coulomb plus linear potential. The $\Pi_u$ and $\Sigma^-_u$ are the $Q\overline{Q}$ potential states with low lying gluonic excitations. Linear plus coulombic potential model is extended in \cite{Nosheen11} for $\Pi_u$ states by fitting the suggested ansatz with lattice data\cite{morningstar03} and the extended model is tested by finding properties of mesons for a variety of $J^{PC}$ states in refs.\cite{Nosheen11,Nosheen14, Nosheen17,Nosheen19}. In this Paper, linear plus coulombic potential model is extended for lowest excited hybrid state, $\Sigma^{-}_{u}$ by fitting the lattice data \cite{morning} with the suggested analytical expression (ansatz). The validity of suggested ansatz is tested by calculating the spectrum of $\Sigma^{-}_{u}$ states and comparing it with lattice results. For this purpose, Born-Oppenheimer formalism and adiabatic approximation is used.  Relativistic corrections in the masses are incorporated through perturbation theory.

Heavy hybrid mesons have been studied using theoretical approaches like lattice QCD  \cite{morning,Bali,2019,morningstar03,Braaten}, constituent gluon model \cite{Iddir07,0611165,0611183},  QCD sum rule \cite{12040049,12094102,12066776,12083273,14106259,14110585,10122614,14037457} and Bethe-Salpeter equation \cite{9802360}.

The paper is organised as: In the section II of this paper, Potential model for $\Sigma^+_g$ state is discussed while the proposed potential model for $\Sigma^-_u$ is defined in section III. The methodology to find radial wave functions, spectrum and radii is explained in section IV , while the discussion on results and concluding remarks are written in section V.

\section*{II. POTENTIAL MODEL FOR $\Sigma^+_g$ STATES}
$\Sigma^+_g$ is the quarkonium state with gound state gluonic field and potential model for this state is defined as \cite{barnes05} :
\begin{equation}
V(r)= \frac{-4\alpha _{s}}{3r}+ br + \frac{32\pi \alpha_s}{9 m_b m_{\overline{b}}} (\frac{\sigma}{\sqrt{\pi}})^3 e^{-\sigma ^{2}r^{2}} \textbf{S}_{b}. \textbf{S}_{\overline{b}} + \frac{4 \alpha _{s}}{m^2_b  r^3} S_T + \frac{1}{m_b^2} \big(\frac{2\alpha _{s}}{r^3}- \frac{b}{2r}\big)\textbf{L}.\textbf{S}.
\end{equation}
Here
$\frac{-4\alpha _{s}}{3r}$ describes coulomb like interaction while linear term $b r$ is due to linear confinement. The term with $\textbf{S}_{b}. \textbf{S}_{\overline{b}}$ is equal to
\begin{equation}
\textbf{S}_{b}.
\textbf{S}_{\overline{b}}=\frac{S(S+1)}{2}-\frac{3}{4}
\end{equation}
$\textbf{L}.\textbf{S}$ describes the spin orbit interactions defined as :
\begin{equation}
\textbf{L}.\textbf{S}=[J(J+1)-L(L+1)-S(S+1)]/2,
\end{equation}
$S_T$ is the tensor operator defined in \cite{barnes05} as :
\begin{equation}
<^{3}L_{J}\mid S_T\mid ^{3}L_{J}>=\Bigg \{
\begin{array}{c}
-\frac{1}{6(2L+3)},J=L+1 \\
+\frac{1}{6},J=L \quad \quad \quad \quad \quad,\\
-\frac{L+1}{6(2L-1)},J=L-1
\end{array}
\end{equation}
Here, $L$ is the relative orbital angular momentum of the quark-antiquark and $S$ is the total spin angular momentum. Spin-orbit and colour tensor terms are equal to zero~\cite{barnes05} for $L=0$. $m_{b}$ is the constituent mass of bottom quark.

\section*{III. POTENTIAL MODEL FOR $\Sigma^-_u$ STATES}

Static potentials for different states of mesons (conventional and hybrid) are computed by lattice simulations. In Figure 3 of ref.\cite{morningstar03}, static potentials of various states are plotted with respect to quark-antiquark seperation. In this paper, potential model (defined in eq. 2) is extended for $\Sigma_u^-$ state by adding the following ansatz:
\begin{equation}
V_\Sigma(r) = A^{\prime} exp (- B^{\prime} r^{P^{\prime}}) + C^{\prime},
\end{equation}
whose parameters are found by fitting it with the lattice data \cite{morningstar03} obtained by taking difference between $\Sigma_{g}^{+}$ and $\Sigma_{u}^{-}$ states. The best fitted values of parameters are:
\begin{equation}
A^{\prime}=11.5917 \textrm{GeV}, \quad B^{\prime}=4.6119, \quad {P^{\prime}}= 0.2810, \quad C^{\prime}=0.9589.
 \end{equation}
With these parameters, dimensionless $\chi^{2}$, defined as :
\begin{equation}
\chi^{2} = \frac{\sum^{^{n}}_{_{i=1}}(\varepsilon_{i} - A exp [- B r_{i}^{2}])^{2}}{\sum^{^{n}}_{_{i=1}} \varepsilon_{i}^2},
\end{equation}
is found to be 0.0000296 for proposed model and lattice data\cite{morningstar03}. Here, $i = 1,2,3,...,n$ is the number of data points. Proposed model ($V_\Sigma(r)$) and lattice data for difference between ($\Sigma_{g}^{+}$) and $\Sigma_{u}^{-}$ potential states are shown in Figure 1.
\begin{figure}
\begin{center}
\epsfig{file=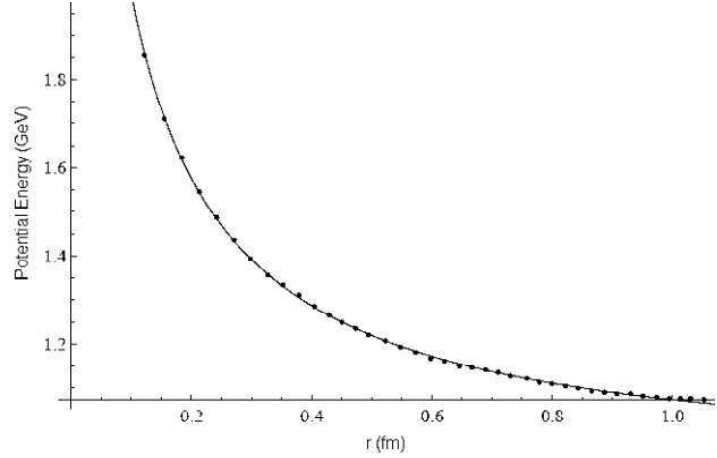,width=0.6\linewidth,clip=}\caption{Curves for potential energy differences between $\Sigma^+_g$ and $\Sigma^-_u$ states. Points represent the potential energy difference taken from ref. \cite{morningstar03} and solid line represents our proposed model}
\end{center}
\end{figure}

\section*{IV. CHARACTERISTICS OF $\Sigma^-_u$ HYBRID BOTTOMONIUM STATES}
\subsection*{1. Radial wave function of $\Sigma^+_g$ and $\Sigma^-_u$ states}

For $\Sigma^+_g$ state, radial Schrodinger equation is written as :
\begin{equation}
U^{\prime \prime }(r)+2\mu \left( E-V(r)-\frac{L(L+1)}{2\mu r^{2}}\right) U(r)=0,
\end{equation}
where V(r) is defined above in eq. (2). Here $U(r)=r R(r)$, where $R(r)$ is the radial wave function. To find numerical solutions of the Schr$\ddot{\textrm{o}}$dinger equation for $\Sigma^+_g$ states, shooting method is used. At small distance (r $\rightarrow$ 0), wave function becomes unstable due to very strong attractive potential. This problem is solved by applying smearing of position co-ordinates by using the method discussed in ref. \cite{godfrey}. To calculate the radial wave functions, parameters $\alpha _{s}= 0.36 $, $b = 0.1340\text{ GeV}^{2}$, $\sigma = 1.34$ GeV, $m_{b} = 4.825$ GeV are taken from ref.\cite{Nosheen17}.

For $\Sigma_u^-$ bottomonium hybrid states, radial Schrodinger equation can be modified as:
\begin{equation}
U^{\prime \prime }(r)+2\mu \left( E-V(r)- A^{\prime} exp(- B^{\prime} r^{P^{\prime}})- C^{\prime}-\frac{L(L+1)-2\Lambda
^{2}+\left\langle J_{g}^{2}\right\rangle}{2\mu r^{2}}\right) U(r)=0,
\end{equation}
Here, $\left\langle J_{g}^{2}\right\rangle $ is the square of gluon angular momentum and $\left\langle J_{g}^{2}\right\rangle =2$~\cite{morning}for $\Sigma^-_u$ state. $\Lambda$ is the projection of gluon angular momentum and $\Lambda =0$ \cite{morning} for $\Sigma^-_u$ state.
Numerical solutions of the Schr$\ddot{\textrm{o}}$dinger equation for $\Sigma^-_u$ states are found by the same method as discussed above and resultant radial wave functions with different $j^{PC}$ are shown in Figure. 2 and Figure 3. The quantum numbers ($L$ and $S$) for these statess are given below in Table 1.

\begin{figure}
\begin{center}
\epsfig{file=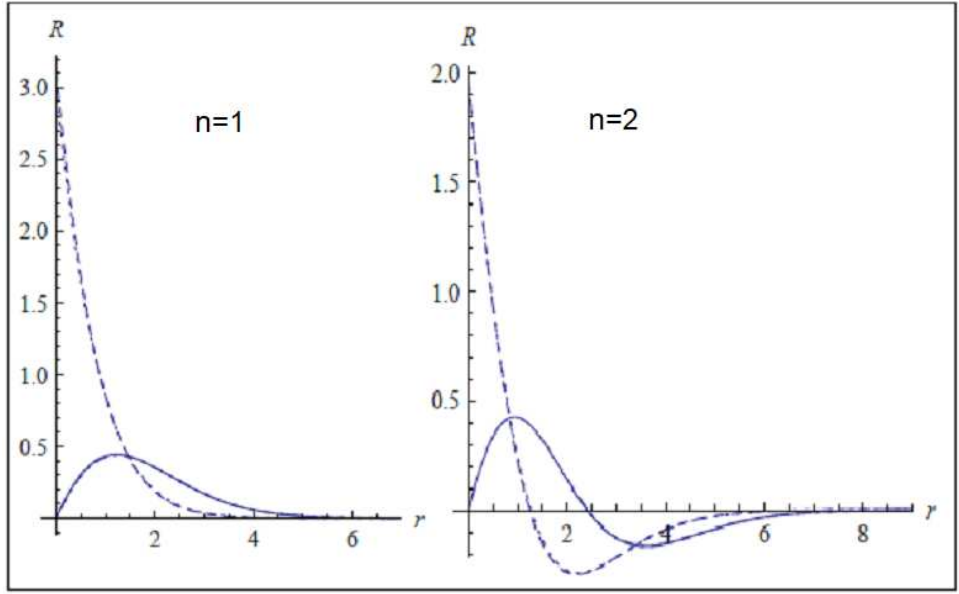,width=0.8\linewidth,clip=}\caption{The radial wave functions for $\Sigma^+_g$ and $\Sigma^-_u$ states for L=0. Solid line curves indicate $\Sigma^-_u$ states and dashed curves are for $\Sigma^+_g$ states. Radial wave functions for $S=0$ and $S=1$ with $L=0$ are almost same in our numerical limits}
\end{center}
\end{figure}

\begin{figure}
\begin{center}
\epsfig{file=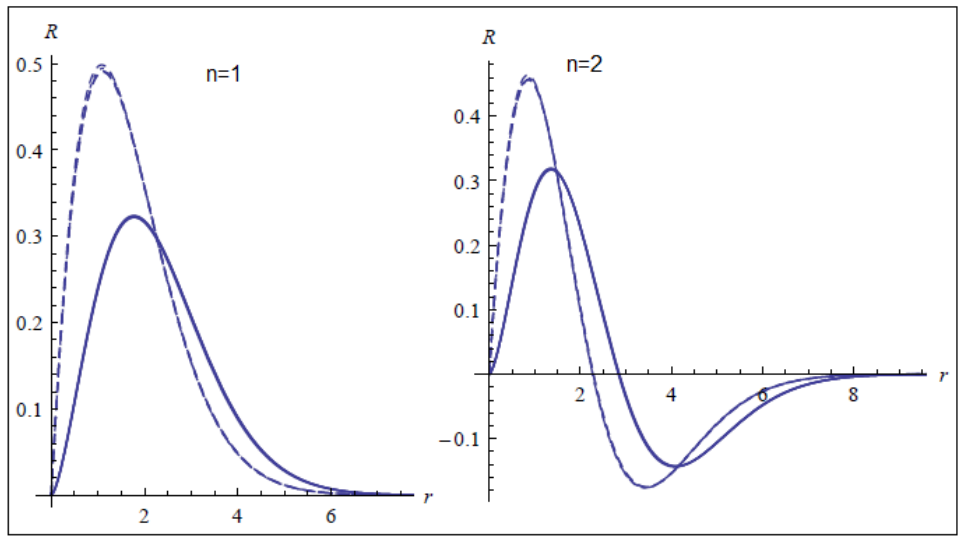,width=0.8\linewidth,clip=}\caption{The radial wave functions for $\Sigma^+_g$ and $\Sigma^-_u$ states for L=1. Solid line curves indicate $\Sigma^-_u$ states and dashed curves are for $\Sigma^+_g$ states. Radial wave functions for $S=0 $ and $S=1$ with $L=1$ are almost same in our numerical limits}
\end{center}
\end{figure}

\subsection*{2. Spectrum of $\Sigma^-_u$ state}

To check the validity of our model, masses of bottomoium mesons are calculated for $\Sigma_u^{-}$ states. To calculate the mass of
 a $b \overline{b}$ state, the constituent quark masses are added to the energy $E$ , i.e;
\begin{equation}
m_{b\bar{b}}=2m_{b}+E,
\end{equation}
The lowest order relativistic correction in mass is incorporated by perturbation theory as adopted in refs. \cite{Nosheen14, Nosheen17} to calculate the spectrum of $\Sigma^+_g$ and $\Pi_u$ state. With relativistic correction, the expression to calculate mass becomes as :
\begin{equation}
m_{b\bar{b}}=2m_{b}+E+\left\langle \Psi \right\vert \left( \frac{-1}{4m_{b}^{3}}\right) p^{4} \left\vert \Psi \right\rangle,
\end{equation}
The best fit values of parameters ($\alpha _{s}= 0.4$, $b=0.11\text{ GeV}^{2}$, $\sigma =1$ GeV, $m_{b}=4.89$ GeV) with relativistic correction are taken from ref.\cite{Nosheen17}. Calculated masses for $\Sigma_u^-$ states with and without relativistic corrections are reported in Table 1.

\subsection*{3. Radii}
The numerically calculated normalized wave functions are used to calculate the root mean square radii. To find the root mean square radii of the gluonic excited $\Sigma_u^-$ bottomonium states, following relation is used:
\begin{equation}
\sqrt{\langle r^{2}\rangle} = \sqrt{\int U^{\star} r^{2} U dr}.\label{P25}
\end{equation}
\\
 \begin{tabular}{|c|}
   \hline
   Table 1
    \\
   \hline
 \end{tabular}
 \\
\section*{V. Discussion and conclusion}

In this paper, potential model for lowest lying $\Sigma_u^-$ hybrid states is proposed whose parameters are found by fitting the model with lattice data\cite{morning, morningstar03}. This model is used to calculate the numerical solutions of Schrodinger equation for $\Sigma_u^-$ states with different $J^{PC}$. In Figures (2-3), normalized radial wave functions of $\Sigma^+_g$ and $\Sigma^-_u$ states are plotted with respect to quark-antiquark seperation $r$. Figures (2-3) show that peaks of radial wave functions are shifted away from origin for gluonic excited states ($\Sigma_u^-$) as compare to gluonic ground states. Figure 2 shows that shape of wave functions is different for $\Sigma_g^+$ and $\Sigma_u^-$ states.

The newly suggested model is used to calculate the masses and radii of the $\Sigma_u^-$ states and results are written in Table 1. Our calculated masses without relativistic corrections are close to the the results given in ref.\cite{2019} as shown in Table 1. In ref. \cite{2019}, spectrum is calculated without including the spin, so the same mass is given for $\eta^h_b$ and $\Upsilon^h_b$. However, our proposed potential model gives distinguished results for $S=0$ and $S=1$. As observed from Table 1, the lowest calculated  mass of the $\Sigma_u^-$ state is calculated to be 10.978 GeV with the incorporation of relativistic corrections in masses. In ref. \cite{morning,morningstar03}, the lowest mass of $\Sigma_u^-$ state is $~11.1$ GeV. This shows that our calculated masses with relativistic corrections are more closer to the masses calculated by lattice simulations \cite{morning, morningstar03} than the nonrelativistic masses.

From Table 1, it is observed that masses and radii are increased by increasing the orbital quantum number (L). The similar behaviour is observed in ref.\cite{Nosheen17} while working on $\Sigma^+_g$ and $\Pi_u$ states of bottomonium meson. Spectrum of $\Sigma^+_g$ state bottomonium mesons is calculated in ref. \cite{Nosheen17} by shooting method and few of the results of ref.\cite{Nosheen17} are shown below in Table 2. The comparison of masses of $\Sigma_g^+$ and $\Sigma_u^-$ states shows that masses and radii of $\Sigma_u^-$ states are greater than $\Sigma_g^+$ states. Overall, we conclude that masses and radii increase towards higher gluonic excitations.
\\
 \begin{tabular}{|c|}
   \hline
   Table 2
    \\
   \hline
 \end{tabular}
\\
Results of calculated radial wave functions, masses and radii can be used to find more properties like decay constant, decay widths and transition rates of gluonic excited $\Sigma_u^-$ states. Overall, we conclude that our extended potential model can be used to study the gluonic excitations in a variety of meson sectors.

\begin{table}[tbp]
\caption{Our calculated masses of $b\overline{b}$ hybrid $\Sigma^-_u$ bottomonium mesons.}
\begin{center}
\tabcolsep=4pt \fontsize{9}{11}\selectfont
\begin{tabular}{|c|c|c|c|c|c|}
\hline
Meson &$J^{PC}$ & \multicolumn{2}{|c|}{calculated mass}& mass & calculated \\ \cline{3-4}
 & & Relativistic & NR  & \cite{2019} & $\sqrt{ \langle r^{2} \rangle}$\\
\hline
& &  & \textrm{GeV} & \textrm{GeV} & fm\\ \hline
$\eta^{h}_{b} (1 ^1S_0)$ &$0^{++}$ & 10.9785 & 10.938 &\multirow{2}{1.4cm}{10.912(3)}&0.4634\\
$\Upsilon^{h} (1 ^3S_1)$ & $1^{+-}$ & 10.9809 & 10.94 & &0.4658 \\ \hline
$\eta^{h}_{b} (2 ^1S_0)$ & $0^{++}$ & 11.2273 &  11.214 &\multirow{2}{1.4cm}{11.192(5)}& 0.7358\\
$\Upsilon^{h} (2 ^3S_1)$ & $1^{+-}$ & 11.2292 & 11.2159 & &0.7368 \\ \hline
$\eta^{h}_{b} (3 ^1S_0)$ & $0^{++}$ & 11.4278 & 11.4411 & &0.9648 \\
$\Upsilon^{h} (3 ^3S_1)$ & $1^{+-}$ & 11.4294 & 11.4429 & &0.9669\\ \hline
$\eta^{h}_{b} (4 ^1S_0)$ & $0^{++}$ & 11.6025 & 11.6413 & &1.1711 \\
$\Upsilon^{h} (4 ^3S_1)$ &$1^{+-}$ & 11.6039 & 11.643 & &1.1729 \\ \hline
$\eta^{h}_{b} (5 ^1S_0)$ & $0^{++}$ & 11.7606 & 11.8238 & & 1.3609 \\
$\Upsilon^{h} (5 ^3S_1)$ & $1^{+-}$ & 11.7618 & 11.8253 & &1.3625 \\ \hline
$\eta^{h}_{b} (6 ^1S_0)$ & $0^{++}$ & 11.9066 & 11.9934 & &1.5384  \\
$\Upsilon^{h} (6 ^3S_1)$ & $1^{+-}$ & 11.9078& 11.9948 & &1.5399  \\ \hline
$h^{h}_{b} (1 ^1P_1) $ & $1^{--}$ & 11.0833 & 11.048 & \multirow{4}{1.4cm}{10.998(4)}&0.5507  \\
$\chi^{h}_{0} (1 ^3P_0)$ &$0^{-+}$ &11.0684 & 11.0424 & &0.5481\\
$\chi^{h}_{1} (1 ^3P_1)$ & $1^{-+}$ & 11.0763 & 11.0477 & & 0.5511\\
$\chi^{h}_{2} (1 ^3P_2)$ & $2^{-+}$ & 11.0872 & 11.0504 & &0.5529\\ \hline
$h^{h}_{b} (2 ^1P_1) $ &$1^{--}$ & 11.3052 & 11.2982 &\multirow{4}{1.4cm}{11.268(6)} &0.8078 \\
$\chi^{h}_{0} (2 ^3P_0)$ & $0^{-+}$ & 11.2926 & 11.2956 & & 0.806 \\
$\chi^{h}_{1} (2 ^3P_1)$ & $1^{-+}$ & 11.3034 & 11.2986 & &0.8085 \\
$\chi^{h}_{2} (2 ^3P_2)$ & $2^{-+}$ & 11.3102 & 11.2998 & &0.8099 \\ \hline
$h^{h}_{b} (3 ^1P_1) $ & $1^{--}$ & 11.4927 & 11.5123 & &1.0297 \\
$\chi^{h}_{0} (3 ^3P_0)$ &$0^{-+}$ & 11.4814 & 11.5109 & &1.0288 \\
$\chi^{h}_{1} (3 ^3P_1)$ &$1^{-+}$ &11.4911 & 11.513 & &1.0306 \\
$\chi^{h}_{2} (3 ^3P_2)$ & $2^{-+}$ & 11.4975 & 11.5137& & 1.0316\\ \hline
$h^{h}_{b} (4 ^1P_1) $ & $1^{--}$ & 11.6596 & 11.7045 & &1.2303 \\
$\chi^{h}_{0} (4 ^3P_0)$ & $0^{-+}$ & 11.6492 & 11.7037 & & 1.2298 \\
$\chi^{h}_{1} (4 ^3P_1)$ &$1^{-+}$ &11.6582 & 11.7053 & &0.2313 \\
$\chi^{h}_{2} (4 ^3P_2)$ & $2^{-+}$ & 11.6641 & 11.7057 & & 1.232 \\ \hline
$h^{h}_{b} (5 ^1P_1) $ & $1^{1--}$ &11.8123 & 11.8814& &1.4159 \\
$\chi^{h}_{0} (5 ^3P_0)$ & $0^{-+}$ & 11.8026 & 11.8809 & & 1.4157\\
$\chi^{h}_{1} (5 ^3P_1)$ & $1^{-+}$ & 11.8109 & 11.8822 & &1.4169\\
$\chi^{h}_{2} (5 ^3P_2)$ & $2^{-+}$ & 11.8165 & 11.8825& &1.4174 \\ \hline
$h^{h}_{b} (6 ^1P_1) $ & $1^{--}$ & 11.9545 & 11.0468 & &1.5901 \\
$\chi^{h}_{0} (6 ^3P_0)$ & $0^{-+}$ & 11.9453 & 11.0466 & & 1.5901\\
$\chi^{h}_{1} (6 ^3P_1)$ & $1^{-+}$ & 11.9531 & 11.0477 & &1.5911 \\
$\chi^{h}_{2} (6 ^3P_2)$ & $2^{-+}$ & 11.9584 & 11.0479 & &1.5916 \\ \hline

$\eta_{b2} (1 ^1D_2)$ & $2^{++}$ & 11.206 & 11.1808 & \multirow{4}{1.4cm}{11.117(4)}& 0.6629 \\
$\Upsilon^{h} (1 ^3D_1)$ & $1^{+-}$ & 11.2004 & 11.1759& & 0.6584 \\
$\Upsilon^{h}_{2} (1 ^3D_2)$ & $2^{+-}$ & 11.2054 & 11.1803 & & 0.6623 \\
$\Upsilon^{h}_{3} (1 ^3D_3)$ & $3^{+-}$ & 11.2088 & 11.1835 & & 0.6657 \\ \hline

$\eta^{h}_{b2} (2 ^1D_2)$ & $2^{++}$ & 11.4041 & 11.4069 & & 0.9045\\
$\Upsilon^{h} (2 ^3D_1)$ & $1^{+-}$ &11.3978 & 11.4042 & & 9017 \\
$\Upsilon^{h}_{2} (2 ^3D_2)$ &$2^{+-}$ & 11.4035 & 11.4068 & & 9044 \\
$\Upsilon^{h}_{3} (2 ^3D_3)$ & $3^{+-}$ & 11.4076 & 11.4085& & 9066 \\ \hline

\end{tabular}%
\end{center}
\end{table}

\begin{table}[tbp]
\caption{Masses of $\Sigma_g^+$ states of bottomonium meson. These results are taken from our earlier work \cite{Nosheen17}}
\begin{center}
\tabcolsep=4pt \fontsize{9}{11}\selectfont
\begin{tabular}{|c|c|c|c|c|c|} \hline
 Meson& Relativistic mass & NR mass & $\sqrt{ \langle r^{2} \rangle}$\\
 & \textrm{GeV} & \textrm{GeV} & $fm$ \\ \hline
$\eta_{b} (1 ^1S_0)$ & 9.4926 & 9.5079 & 0.2265 \\
 $\Upsilon (1 ^3S_1)$ & 9.5098 & 9.5299 & 0.2328 \\ \hline
$\eta_{b} (2 ^1S_0)$ & 10.0132 & 10.0041 & 0.5408\\
$\Upsilon (2 ^3S_1)$ & 10.0169 & 10.0101 & 0.5448 \\ \hline

$h_{b} (1 ^1P_1) $ & 9.9672 & 9.9279 & 0.4347\\
$\chi_{0} (1 ^3P_0)$ & 9.8510 & 9.9232 & 0.4375\\
$\chi_{1} (1 ^3P_1)$ & 9.9612 & 9.9295 & 0.4379\\
$\chi_{2} (1 ^3P_2)$ & 9.9826 & 9.9326 & 0.4375\\ \hline

$\eta_{b2} (1 ^1D_2)$ & 10.1661 & 10.1355 & 0.5933\\
$\Upsilon (1 ^3D_1)$ & 10.1548 & 10.1299 & 0.5930\\
$\Upsilon_{2} (1 ^3D_2)$ & 10.1649 & 10.1351 & 0.5939\\
$\Upsilon_{3} (1 ^3D_3)$ & 10.1772 & 10.1389 & 0.5942\\ \hline
\end{tabular}
\end{center}
\end{table}

\end{document}